\documentclass[twocolumn]{jpsj2} 
\title{Pressure Evolution of the Ferromagnetic and Field Re-entrant Superconductivity in URhGe}

\author{Atsushi \textsc{Miyake}\thanks{E-mail address: amiyake@cqst.osaka-u.ac.jp,
Present address:  KYOKUGEN, Osaka University, Toyonaka, Osaka 560-8531, Japan. }, Dai \textsc{Aoki}, and Jacques \textsc{Flouquet}}

\inst{%
INAC/SPSMS, CEA-Grenoble, 17 rue des Martyrs, 38054 Grenoble, France 
}

\abst{
Fine pressure ($P$) and magnetic field ($H$) tuning on the ferromagnetic superconductor URhGe are reported in order to clarify the interplay between the mass enhancement, low field superconductivity (SC) and field reentrant superconductivity (RSC) by electrical resistivity measurements.
With increasing $P$, the transition temperature and the upper critical field of the low field SC decrease slightly, while the RSC dome drastically shifts to higher fields and shrinks.
 The spin reorientation field $H_{\rm R}$ also increases.
At a pressure $P\sim 1.8$~GPa, the RSC has collapsed while the low field SC persists and may disappear only above 4~GPa.
Via careful $(P, H)$ studies of the inelastic $T^2$ resistivity term, it is demonstrated that this drastic change is directly related with the $P$ dependence of the effective mass which determines the critical field of the low field SC and RSC on the basis of triplet SC without Pauli limiting field.} 

\kword{URhGe, ferromagnetic superconductor, field reentrant superconductivity, pressure, effective mass, electrical resistivity }

\begin{document}
\maketitle

The discoveries of superconductivity (SC) inside the itinerant ferromagnetic (FM) phase of different uranium intermetallic compounds \cite{saxena2000sbi, Aoki2001, Akazawa2004, Huy2007} were spectacular events in the field of unconventional superconductivity even though this possibility was predicted a long time ago \cite{Fay1980}.
The main reason given for this occurrence is that a triplet pairing is presumably generated by FM spin fluctuation.
For the two examples UGe$_2$ \cite{sheikin2001rcs} and URhGe \cite{hardy2005pws} it seems established that equal spin pairing is realized from the temperature dependence of upper critical field $H_{c2}(T)$.

In a simple picture, a magnetically mediated SC is believed to be stuck to the so-called quantum critical point (QCP) where the magnetic ordering temperature collapses as a function of an external parameter such as pressure \cite{mathur1998mms}.
For the specific case of the itinerant FM systems, it is well established theoretically and experimentally that the nature of the quantum phase transition (QPT) at $P_c$, where the ground state switches from FM to paramagnetic (PM), is not second order but first order \cite{saxena2000sbi, Huxley2001}.
Furthermore, it has been stressed that inducing a finite polarization by an external field, a tricritical point (TCP) and two quantum critical points in the vicinity of $P_c$ appear with metamagnetic phenomena \cite{Belitz2005}.   

To illustrate these effects, we present here the results on the ($T, P, H$) variation of FM and SC of URhGe including the effect on the field reentrant superconductivity (RSC) \cite{Levy2005}.
Applying pressure on URhGe will allow us to go deeper in the FM domain, thus moving away from the QCP and TCP, as the Curie temperature $T_{\rm Curie}$ increases under pressure \cite{hardy2005ptp}.


URhGe with an orthorhombic crystal structure orders ferromagnetically below $T_{\rm Curie}$= 9.5~K and becomes SC below $T_{\rm sc} \sim 0.25$~K at zero field and ambient pressure \cite{Aoki2001}. 
When $H$ is swept along the $b$-axis, a jump of the magnetization occurs along the $b$-axis associated with the collapse of magnetization component along the $c$-axis for $H > H_{\rm R}$=~12~T at which the induced magnetization $M_{\rm R}$ reaches a critical value of $\sim 0.3~\mu_{\rm B}$, comparable to the full FM ordered moment $M_0\sim 0.4~\mu_{\rm B}$ at zero field along the easy $c$-axis \cite{Levy2005}.
The reorientation field $H_{\rm R}$ is marked by a clear resistivity peak \cite{Levy2005}.
The RSC occurs between $H_2 = 8$~T and $H_3 = 12.7$~T (slightly higher than $H_{\rm R}$ for a perfect field alignment along the $b$-axis) \cite{Levy2005}.
The transition temperature $T_{\rm sc}$ of RSC is larger than that at zero field \cite{Levy2005}.
In addition, the RSC phase persists in a wide range of the field direction from $b$- to $a$-axis \cite{levy2007aeu}, while the change of the applied field direction from the $b$- to $c$-axis shifts $T_{\rm RSC}$ ($H_{\rm R}$) to lower temperature (higher field) and RSC disappears around 7$^{\circ }$ tilted from the $b$-axis \cite{Levy2005}. 
This change near QCP will have consequences on the spin fluctuation i.e. the effective mass $m^{\ast}_H$ \cite{Miyake2008}.
Similarly to heavy fermion system cases, the metamagnetic field $H_{\rm R}$ is suspected to vary with pressure as $M_{\rm R} = \chi H_{\rm R}$ with the Pauli susceptibility $\chi \propto m^{\ast}$.
In URhGe, it is expected that $H_{\rm R}$ will increase with decreasing $m^{\ast}$ by applying pressure, assuming $P$-invariable $M_{\rm R}$.


Our previous investigation on the strong sample quality dependence of $T_{\rm sc}$ revealed that both the low-field SC and RSC are unconventional \cite{Miyake2008}.
On the basis of McMillan-type formula in the itinerant FM phase \cite{Fay1980, Kirkpatrick2001}, we could reproduce the low field SC but also the RSC as a function of the effective mass $m^{\ast}_H$ derived from the $T^2$-resistivity coefficient $A$. 
It has been concluded that the field dressing of the effective mass $m^{\ast \ast}_H$, which is related with the band mass $m_{\rm B}$ as $m^{\ast}_{H}=m_{\rm B}+m^{\ast\ast}_H$, through the FM fluctuation and the spin reorientation near $H_{\rm R}$ plays an important role in the occurrence of the two superconducting states \cite{Miyake2008}.

We also proposed the relation between the effective mass and $T_{\rm sc}$ through the effective Gr$\ddot {\rm u}$neisen parameter, $\Omega_{T_{\rm sc}} = \frac{1}{\lambda} (\Omega_{T_{\rm B}} - \Omega_{T^{\ast\ast}} )$, where the characteristic temperatures $T_{\rm B}$ and $T^{\ast\ast}$ are inversely proportional to $m_{\rm B}$ and $m^{\ast\ast}$, respectively, and the coupling constant $\lambda = m^{\ast\ast}/m^{\ast}$ \cite{Miyake2008}. 
Comparing the pressure dependence of $T_{\rm sc}$ to heavy fermions and high $T_{\rm sc}$ cuprates, the change of $m_{\rm B}$ is the dominant mechanism for $T_{\rm sc}$ in the cuprate: $\Omega_{T_{\rm sc}} \sim \Omega_{T_{\rm B}}$ \cite{Nakamura1996}.
On the other hand, for the superconducting uranium compounds like UPt$_3$, URu$_2$Si$_2$ and UBe$_{13}$ $T_{\rm sc}$ varies with $m^{\ast\ast}$: $\Omega_{T_{\rm sc}} \sim -\Omega_{T^{\ast\ast}}$ \cite{Flouquet1991}. 
URhGe can also be classified with the latter examples.

A high quality single crystal of URhGe was prepared by the Czochralski method and the details are already given in ref.~\citen{Miyake2008}.
The residual resistivity ratio ($\rho_{\rm RT}/\rho_0$) used here is roughly 50, which is the best sample that we have studied.
We employed a piston-cylinder type pressure cell with Daphne 7373 oil as a pressure-transmitting medium.
The applied pressure was determined by the pressure dependence of $T_{\rm sc}$ of Pb.
The electrical resistivity for the current along the $a$-axis was measured by the four probe AC method at temperatures down to 80~mK and at magnetic fields along the $b$-axis up to 16~T.
Two different $P$ runs were performed.
In the run 1, $P$ was increased from 0 to $\sim$ 0.9~GPa, in the run 2 from 0.6 to 1.8~GPa.
Due to the high sensitivity of RSC to misalignment, renormalized plots will often be drawn.

\begin{figure}[tb]
\begin{center}
\includegraphics[width=0.9 \hsize,clip]{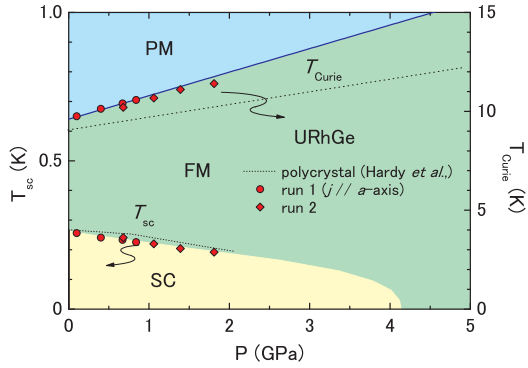}
\caption{\label{PTPD}
(Color online) Pressure dependence of $T_{\rm sc}$ (left scale) and $T_{\rm Curie}$ (right scale) of URhGe.
The dotted lines are previous results on polycrystalline sample taken from ref.~\citen{hardy2005ptp}.
The solid line indicates the pressure dependence of $T_{\rm Curie}$, $dT_{\rm Curie}/ dP \sim$~1.2~K/GPa expected from the Ehrenfest relation \cite{sakarya2003dsf}. 
 }
\end{center}
\end{figure}

As reported \cite{hardy2005ptp}, $T_{\rm Curie}$ ($T_{\rm sc}$) increases (decreases) with pressure as shown in Fig.~\ref{PTPD}.
Compared to the results of polycrystalline sample, the present results are more strongly dependent on pressure.
Furthermore, the agreement of $T_{\rm Curie} (P)$ with the expectation from the Ehrenfest relation reported previously is excellent \cite{sakarya2003dsf}. 
For all the measured pressures up to 1.8~GPa and fields up to 16~T, no deviation from Fermi-liquid law in the resistivity ($\rho \propto T^2$) was observed at low temperature below $T =1.2~{\rm K}$. 
As shown later, the decrease of the $A$-coefficient with pressure is in agreement with the proposed relation $\Omega_{T_{\rm sc}} \sim -\Omega_{T^{\ast\ast}}$.

\begin{figure}[bt]
\begin{center}
\includegraphics[width=0.8 \hsize,clip]{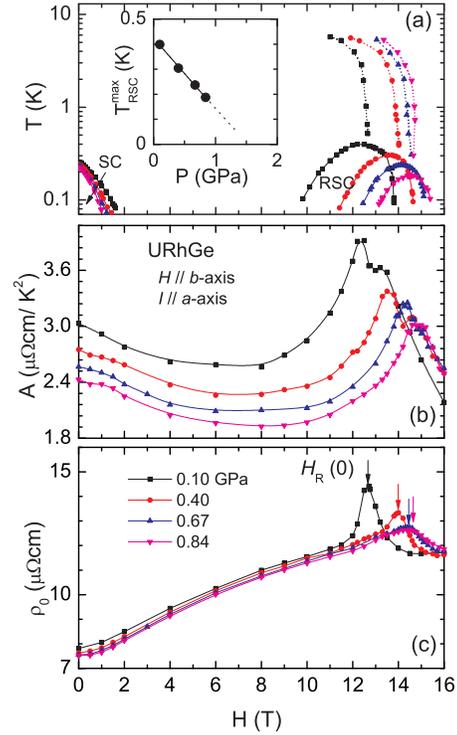}
\caption{\label{HTPD} 
(Color online)
(a)Temperature-field phase diagram of the low field superconductivity (SC), the reentrant SC (RSC), and the reorientation field $H_{\rm R}$ for $H || b$ in URhGe under pressures. 
The inset of panel(a) is the pressure dependence of maximum value of $T_{\rm sc}$ for RSC, $T_{_{\rm RSC}}^{\rm max}$.
The field variation of the coefficient of $T^2$-term of the resistivity $A$ (b) and the residual resistivity $\rho_0$ (c) under several pressures.
The arrows in panel (c) indicate the zero temperature extrapolation of $H_{\rm R}$.}
\end{center}
\end{figure}

\begin{figure}[tb]
\begin{center}
\includegraphics[width=0.8 \hsize,clip]{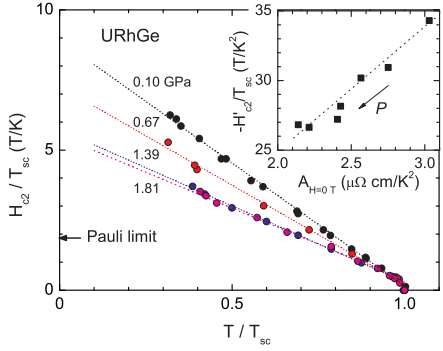}
\caption{\label{Hc}
(Color online)
Upper critical field $H_{c2}$ of URhGe normalized to the $T_{\rm sc}(H=0~{\rm T})$ as a function of the reduced temperature of $T/T_{\rm sc}$ at pressures.
Pauli limiting field at 0~K is also indicated with an arrow.
The inset shows the comparison between the $A$ coefficient and the $-H'_{c2}/T_{\rm sc}\propto (m^{\ast})^2$ obtained by linear fitting shown the dotted line in the main figure. }
\end{center}
\end{figure}

\begin{figure}[tb]
\begin{center}
\includegraphics[width=0.8 \hsize,clip]{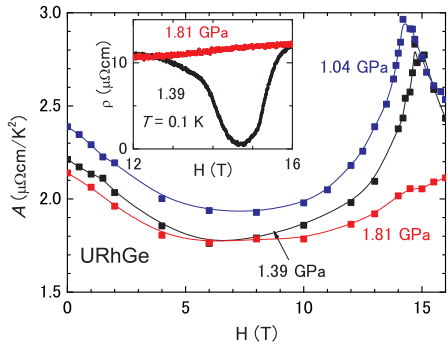}
\caption{\label{AH}
(Color online)
Field evolution of  the $A$ coefficient under pressures (run 2).
The strong suppression of a peak $A$ at 1.81~GPa is in excellent agreement with the disappearance of RSC shown in the inset.
The inset presents the field dependence of resistivity at $T=$0.1~K and for $P=$~1.39 and 1.81~GPa.}
\end{center}
\end{figure}

\begin{figure}[tb]
\begin{center}
\includegraphics[width=0.8 \hsize,clip]{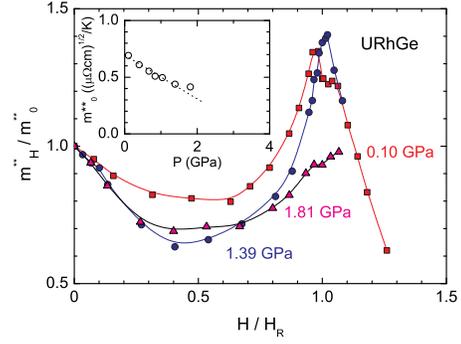}
\caption{\label{mH}
(Color online)
Field evolution of  the correlation effective mass $m^{\ast\ast}_{H}$ at 0.1~GPa (run 1), 1.39~GPa and 1.81~GPa (run 2).
The field and $m^{\ast\ast}_{H}$ is normalized by $H_{\rm R}$ and $m^{\ast\ast}_0$ respectively.
The inset shows pressure dependence of $m^{\ast\ast}_0$.
The low field SC will be suppressed when $m^{\ast\ast}_0$ becomes zero in our model.}
\end{center}
\end{figure}

The obtained ($H,~T$)-phase diagrams at several pressures of run 1 are presented in Fig.~\ref{HTPD}(a). 
$T_{\rm sc}$ was determined as the half drop of the resistivity as a function of field and temperature.
Both SC domes become smaller with pressure.
The RSC phase strongly shifts towards higher fields, together with the reorientation field $H_{\rm R}$.
Moreover, the maximum value of $T_{\rm sc}$ for the RSC phase, $T^{\rm max}_{_{\rm RSC}}$ is more strongly suppressed than $T_{\rm sc}$ at zero field.
As shown in the inset of Fig.~\ref{HTPD}(a), a linear $P$ extrapolation leads to predict that RSC will disappear at $P_{\rm RSC} \sim$~1.5~GPa.
In the $\rho(H)$ curves, a sharp peaked anomaly corresponding to the spin reorientation at $H_{\rm R}$ appears above $T_{\rm sc}$ and is also clearly visible in the zero temperature extrapolation $\rho_0(H)$ (see Fig~\ref{HTPD}(c)).  
As clearly seen, the spin reorientation field $H_{\rm R}$ is connected strongly to the RSC phase.

According to our simple model for $T_{\rm sc}$, $T_{\rm sc}$ varies with $m^{\ast\ast}_H$. 
In Fig.~\ref{PTPD}, $T_{\rm sc}$ at zero field decreases with increasing pressure, and hence, the decrease of the effective mass ($m^{\ast}_0$) is expected.  
As seen in Fig.~\ref{Hc}, the reduced scale $H_{c2}$ is proportional to $T$ down to at least $T\sim 0.3T_{\rm sc}$,
and the value $H_{c2}/T_{\rm sc}$ at $T=0$~K is much larger than the Pauli limit ($H_{\rm c2}(0)/T_{\rm sc} = 1.86$~T/K) assuming the BCS weak coupling model.
This indicates that $H_{c2}$ is governed by the orbital limit without Pauli limitation, which is consistent with the spin triplet SC for equal spin pairing. 
In general, the effective mass can be derived from the initial slope of $H_{c2}(T)$-curve at $T_{\rm sc}$, i.e. $-H'_{c2} = -(dH_{c2}/dT)_{T=T_{\rm sc}}\propto (m^{\ast})^2 T_{\rm sc}$.
Because of the SC in the FM phase, the strong initial slope close to $T_{\rm sc}$ arises from the magnetization due to FM ordering and is not taken into account here.
The effective mass is reduced by $P$ in excellent agreement with both the pressure dependences of $A$ at $H=$0~T and $-H_{c2}'/T_{\rm sc}$ as shown in the inset of Fig.~\ref{Hc}.
These facts confirm the reliability of the qualitative determination of the effective mass from $\sqrt A$ based on the Kadowaki-Woods relation, and allow us to analyze the relation between the effective mass and $T_{\rm sc}$.

Fig.~\ref{HTPD}(b) and (c) represent the field variation of $A$ and $\rho_0$ under pressure obtained by a least square fitting with the Fermi-liquid law $\rho = \rho_0 + AT^2$ in the run~1.
The $A$ values show clear peaks at the SC phases and decrease with increasing pressure.
$\rho_0$ at low field extrapolated to 0~K from the normal FM phase is almost invariant but decreases slightly with pressure.
A clear peak of $\rho_0$ corresponding to $H_{\rm R}$ shifts higher field with some broadening as a function of pressure, although the $A(H)$ curve still shows a sharp peak at slightly lower field than $H_{\rm R} (T = 0)$.
Pressure changes the shape of the field dependence of $\rho_0$ from a clear sharp peak to a shoulder-like anomaly.
It suggests an evolution in the process of magnetization reorientation under pressure.
Looking below to the field variation of the inelastic $T^2$-term this drastic regime change will be confirmed.
The similar tendency was observed with slightly rotating the applied field from the $b$- to $c$-axis \cite{Levy2005}. 
Hence, pressure seems to act on the field evolution of magnetization quite similarly to a field misalignment.

Fig.~\ref{AH} represents the field variation of the $A$ coefficient at higher pressure.
Up to 1.39~GPa, the relative $H$ enhancement of $A(H_{\rm R})/A(0)$ appears almost $P$ invariant while $A(0)$ decreases.
At $P=$~1.81~GPa, the sharpness of the field enhancement of $A(H)$ on approaching $H_{\rm R}$ is replaced by a broad feature.
As shown in the inset of the Fig.~\ref{AH}, no RSC can be detected at $P = 1.8$~GPa.
Even at high fields above 16~T, RSC cannot be expected, as seen in the inset of Fig.~2(a).

\begin{figure}[tb]
\begin{center}
\includegraphics[width=0.8 \hsize,clip]{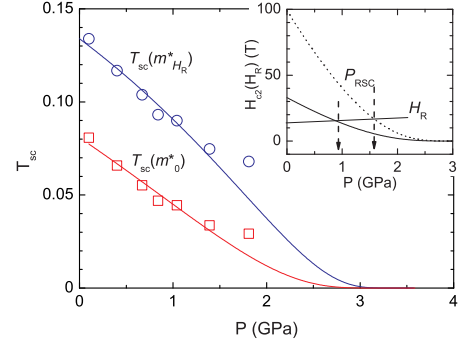}
\caption{\label{CalTc}
(Color online)
Calculated $T_{\rm sc} = \exp{(-m^{\ast}_H/m^{\ast\ast}_H)}$ at $H = 0$ (open squares) and $H_{\rm R}$ (open circles) as a function of pressure.
The ratio of $m^{\ast\ast}_{H_{\rm R}}/m^{\ast\ast}_{0}= 1.5$ is assumed to be independent of pressure, which is not far from experimental results shown in Fig.~\ref{mH}. 
The solid lines are fitting results assuming $m^{\ast\ast}_0$ linearly decreases with pressure as shown in the inset of Fig.~\ref{mH}.
The inset shows the calculated $H_{c2}(m^{\ast}_{H_{\rm R}})= K[m^{\ast}_{H_{\rm R}}T_{\rm sc} (m^{\ast}_{H_{\rm R}})]^2$, where $K$ being a scaling factor determined at 0.1~GPa (solid line) and 1.4~GPa (dotted line).
}
\end{center}
\end{figure}

Assuming that the pressure and field invariant renormalized band mass $m_{\rm B}$ corresponds to $A_{\rm B}= 1.1~\mu\Omega$cmK$^{-2}$, one can derive the field dependence of $m^{\ast\ast}_H = m^{\ast}_H - m_{\rm B}$.
Figure~\ref{mH} represents the renormalized enhancement of $m^{\ast\ast}_H$ by comparison to the zero field value $m^{\ast\ast}_0$ as a function of $H/H_{\rm R}$.
Up to $P\sim 1.4~{\rm GPa}$, the ratio $m^{\ast\ast}_{H=H_{\rm R}}/m^{\ast\ast}_0$ seems to be constant at $\sim$1.5, while $m^{\ast\ast}_0$ decreases with pressure.
Taking the same crude hypothesis as that in ref.~\citen{Miyake2008}, one can calculate for $H=0$ what will be $T_{\rm sc}$ for mass enhancement $m^{\ast\ast}_0$ and $m^{\ast\ast}_{H_{\rm R}}$ assuming $m^{\ast\ast}_{H_{\rm R}}/m^{\ast\ast}_0 = 1.5$.
The concomitant $H$ increases of $m^{\ast}_{H_{\rm R}}$ and $T_{\rm sc}(H_{\rm R})$ lead to a superconducting coherence length at $H_{\rm R}$ shorter than at $H=0$.
The decrease of the electronic mean free path ($\propto 1/\rho_0$) at $H_{\rm R}$ observed on Fig.~\ref{HTPD} (c) gives quite similar criteria for SC between $H=0$ and $H_{\rm R}$ to fulfill the clean limit condition.
Thus at low pressure RSC is as robust as low field SC.
The disappearance of RSC is governed by the other condition that $H_{c2}(m^{\ast}_{H_{\rm R}}, T\rightarrow  0~{\rm K})= K [m^{\ast}_{H_{\rm R}}T_{\rm sc}(m^{\ast}_{H_{\rm R}})]^2$ is greater than $H_{\rm R}$.
That is drawn in the inset of Fig.~\ref{CalTc} with the options to fit the scaling constant $K$ by the values of $H_3$ either at $P= 0.1$ and 1.4~GPa.
The predicted collapse pressure of RSC agrees with the expected $P_{\rm RSC}\sim 1.5$~GPa (Fig.~\ref{HTPD}) and the observed lack of RSC at 1.8~GPa in the resistivity (the inset of Fig.~\ref{AH}).  
Entering more deeply in the FM domain (presumably escaping from TCP), a change of regime clearly occurs near $P=1.8$~GPa: the sharp peak structure in the mass enhancement at $H_{\rm R}$ is smeared out (Fig.~\ref{mH}).

The initial weak $P$ dependence of $m^{\ast\ast}_{H_{\rm R}}/m^{\ast\ast}_0$ of URhGe (Fig.~\ref{mH}) is remarkable.
This persistence in the field enhancement is reminiscent of effects observed in the pressure study of the pseudo-metamagnetism of the heavy fermion compound CeRu$_2$Si$_2$ at $H_{\rm M}$\cite{Mignot1989}.
In CeRu$_2$Si$_2$ despite a huge pressure decrease of $A(0)$, there is an initial $P$ quasi-invariance of $A(H_{\rm M})$ as if the magnetic field drives the system near the same quantum critical fluctuations with a finite effective value of $m^{\ast\ast}(H_{\rm M})$.
Of course at very high pressure, $A(H_{\rm M})/A(0)$ will vanish with the simultaneous collapse of FM and antiferromagnetic fluctuations. 

To summarize, for superconductivity the collapse of RSC is governed by the condition $H_{c2}(m^{\ast})$ higher than the applied $H$.
Concerning the normal FM phase, around $P\sim$ 1.8~GPa, there is a change in the regime of the field mass enhancement at $H_{\rm R}$, i.e. $H$ spin fluctuation dynamics as detected by the measurement of $A(H)$, corresponding to the collapse of RSC.
Let us point out that it coincides with the change of $A$ in the pressure dependence of $m^{\ast\ast}_0$ observed near 1.4~GPa.
The smearing in the $H$ enhancement at 1.8~GPa indicates that URhGe is near its TCP at ambient pressure, and gets close to the regime of conventional itinerant ferromagnetism with pressure.

\section*{Acknowledgment}
We thank D. Braithwaite for help in manuscript preparation.
This work was financially supported by French Agence Nationale de la Recherche through Contracts ANR-06-BLAN-0220 ECCE and  ANR-07-CEXC-004-01 CORMAT.


\end{document}